\documentclass{aau}
\usepackage{graphicx}
\def\I{\mbox{IRAS~23151+5912}}
\newcommand{\beq}{\begin{equation}}
\newcommand{\eeq}{\end{equation}}
\newcommand{\D}{\displaystyle}
\newcommand{\rrho}{\varrho}
\begin{document}
\title{Bispectrum speckle interferometry of the massive protostellar
outflow source IRAS~23151+5912\thanks{Based
on data collected at the MMT 6.5~m telescope.}}

\author{G.~Weigelt\inst{1},
H.~Beuther\inst{2},
K.-H.~Hofmann\inst{1}, M.R.~Meyer\inst{3}, Th.~Preibisch\inst{1},
D.~Schertl\inst{1},
M.D.~Smith\inst{4}, \and E.T.~Young\inst{3}}

\offprints{Gerd Weigelt}

\institute{Max-Planck-Institut f\"ur Radioastronomie,
 Auf dem H\"ugel 69, D--53121 Bonn, Germany
\and
Harvard-Smithsonian Center for Astrophysics, 60 Garden Street, Cambridge, MA 02138, USA
\and Steward Observatory, University of Arizona, 933 North Cherry Avenue, Tucson, AZ 85721, USA
\and  Armagh Observatory, College Hill, Armagh BT61 9DG, Northern Ireland
}

   \date{Submitted: ; accepted: }

   \abstract{
We present  bispectrum speckle interferometry
of the massive protostellar object \I\, in the near-infrared $K'$ band.
The reconstructed image shows the  diffuse nebulosity 
north-east of two point-like sources in unprecedented detail.
The comparison of our near-infrared image with mm continuum and CO 
molecular line maps shows
that the brighter of the two point sources lies near the center of the 
mm peak, indicating that it is a high-mass protostar.
The nebulosity coincides with the blue-shifted molecular
outflow component. The most prominent feature in the nebulosity is
a bow-shock-like arc.
We  assume that this feature is associated with a precessing jet
which has created an inward-pointed cone in the swept-up material.
We present numerical jet simulations that reproduce this and several
other features observed in our speckle image of the nebulosity.
Our data also reveal a  linear structure connecting
the central point source to the extended
diffuse nebulosity. This feature may represent the innermost part
of a jet that drives the
strong molecular outflow (PA~$\sim 80\degr$) from \I. 
With the aid of radiative transfer calculations, we demonstrate 
that, in general, the observed inner 
structures of the circumstellar material surrounding high-mass stars 
are strongly influenced by the
orientation and symmetry of the bipolar cavity. 
   }
\authorrunning{Weigelt et al.} \titlerunning{Infrared speckle imaging of \I}
   \maketitle
%

\section{Introduction}
The formation of massive stars is usually associated with strong
outflow activity. In fact, many massive protostars have been
identified by their massive molecular outflows observed at radio wavelengths.
These outflows seem to be an essential ingredient of the star formation
process (for a review see, e.g., Reipurth \& Bally 2001)
as they are believed to contribute to
the removal of excess angular momentum from accreted matter and to
disperse infalling circumstellar envelopes.
Despite their key role in star formation, the
origin, acceleration, and collimation of the flows are still poorly
understood, mainly because these processes are believed to occur on 
very small spatial scales. 
This clearly demonstrates the need 
for high spatial-resolution studies in order to reach a better understanding of 
the star-formation process.

For low-mass young stellar objects (YSOs),
many observations have indicated that the molecular outflows are momentum-driven
by protostellar jets, which entrain the surrounding material.
In the formation of high-mass stars,
it is unclear whether we simply see
 a scaled-up version of the processes working in low-mass objects
or whether there are fundamental differences between the high- and low-mass
star formation regimes. It has often been argued  that
the outflows from high-mass protostars generally display a considerably
more complex structure and are less collimated
than outflows from low-mass YSOs,
suggesting fundamental differences. Alternative 
outflow driving mechanisms for massive YSOs
include the deflection of accreted material
(Churchwell 2000) and the strong winds of hot, massive stars
(e.g.~Devine et al.~1999).

Beuther et al.~(2002a) studied  the bipolar outflows 
in a large sample of high-mass star-forming regions.
They found generally continuous correlations
of the outflow properties from 
the low-mass to the high-mass regime, suggesting a common mechanism for the 
origin of the outflows for all masses.
Their mm data are consistent with massive outflows being as
collimated as their low-mass counterparts.

High spatial-resolution imaging of the inner circumstellar 
environment in YSOs can shed
light on the origin of the flows and the 
mechanism responsible for the initial collimation of the beams.
Studies with the HST, adaptive optics,  
and infrared long-baseline interferometry
have recently yielded important information
on the inner circumstellar environment of many YSOs 
(e.g.~Stapelfeldt et al.~1998; Brandner et al.~2000; 
Millan-Gabet \& Monnier 2002; Leinert et al.~2004;
Eisner et al.~2005; Akeson et al.~2005)
and have revealed spectacular manifestations of the interaction of the jets
and outflows with the surrounding material (e.g.~Dougados et al.~2000).
Bispectrum speckle interferometry
has revealed inner circumstellar structures of very complex morphology
around several massive and intermediate-mass YSOs
(e.g.,~Schertl et al.~2000; Weigelt et al.~2002ab; Preibisch et al.~2001, 2002,
2003; Hofmann et al.~2004).
These results were the motivation for our speckle interferometric study of the
YSO \I\, presented in this paper.
If we can distinguish infalling envelope material from outflows surrounding
massive stars, we will gain insight into: 1) whether the relationship between
mass accretion and mass loss for low-mass stars  also applies to
higher mass
protostars; 2) the injection of kinetic energy into molecular cloud cores
from the formation of massive stars;
and 3) the effects that outflows might have on the emergent initial mass
function in massive
star-forming regions.  Our long-term program attempts to answer these questions
with a systematic study of massive protostars at high angular resolution
utilizing the techniques of bispectrum speckle interferometry.
\medskip

The infrared source \I\ is located in a rather isolated molecular cloud 
in the Cepheus region to the south of the HII region Sh2-157.
The IRAS fluxes show a strongly rising broad-band spectrum
with a peak at $60\,\mu$m, demonstrating the presence of large amounts
of warm dust.
The ISO SWS spectrum rises very steeply between $4\,\mu$m 
 and $45\,\mu$m
and shows a deep and broad silicate absorption feature around
$10\,\mu$m.
The ISOPHOT spectrum shows another strong absorption 
feature at $3\,\mu$m, which can be explained by ice mantles on the
dust grains. These two absorption features suggest that 
the central source is deeply embedded in dense circumstellar material.

While \I\ is completely invisible at optical wavelengths, it is a bright
object in the near-infrared $K$ band.
The $K$-band image by Hodapp (1994;  seeing-limited resolution of $\sim 1''$) 
shows two point sources and a diffuse nebula.
A near-infrared imaging ($\sim 1.8''$ resolution) and spectroscopic study of \I\
and its surroundings was recently presented by Chen \& Yao (2004).

A bipolar molecular outflow from \I\ was detected in the CO observation
by Wouterloot et al.~(1989).
\I\ is part of a large sample of 69 very young high-mass
star-forming regions intensely studied by Sridharan et al. (2002) and
Beuther et al.~(2002a,b,c). 
We will use the source parameters listed in
Beuther et al.~(2002b). The source is at a distance of $\sim
5.7$\,kpc with a bolometric luminosity of
$10^5\,L_{\odot}$, suggesting a mass of about $25\,M_\odot$ for the 
YSO. 
The source exhibits a massive bipolar outflow in the east-west direction;
the direction of the line connecting the peaks of the red- and blue-shifted
CO maps is along position angle PA = $79\degr$.
The masses of the red and blue outflow lobes are $\sim 8\,M_\odot$ and
$\sim 13\,M_\odot$, respectively.
The size of the outflow system is 0.55~pc and the collimation factor
(i.e.~the ratio of the length to the width of the flow) is 1.2.
Note that this relatively small collimation factor does not necessarily 
imply that the outflow is poorly collimated, since this value
strongly depends on the beam size
and the inclination angle (which is probably quite large).
The dynamical age of the outflow is estimated to be only $20\,000$~years.

Wouterloot \& Walmsley (1986), Scalise et al.~(1989), Felli et al.~(1992),
Tofani et al.~(1995), and Beuther et al.(2002c) detected H$_2$O maser emission in \I. 
The presence of these masers 
is an additional signpost of the youth of this object. 
Furthermore, the non-detection of \I\ at 3.6\,cm with
the VLA at a 1\,mJy sensitivity limit indicates
that the object is in a very early
evolutionary stage prior to forming a significant ultracompact H\,{\sc ii}
region.


\section{Observations and data analysis}

The speckle interferograms of \I\,were recorded on 20 December 2004
with the 6.5~m MMT in Arizona.
The detector of our speckle camera was a Rockwell HAWAII array
(only one $512\times 512$ quadrant was used).
The size of one pixel corresponds to 27.0\,mas on the sky.
A K-band filter with central wavelength 2115\,nm and bandwidth 214\,nm
was used. This filter bandwidth includes the 
1-0~S(1) emission line of molecular hydrogen at 2.12\,$\mu$m and 
and the Br\,$\gamma$ line.
The speckle interferograms of the unresolved star HD~240212 were used for the
compensation of the atmospheric speckle transfer function.
The exposure time per frame was 590~ms.
Our data set consists of 350 speckle interferograms of \I\ and
150 speckle interferograms of the unresolved reference star HD~240212.
The  field-of-view was $11.3'' \times 11.3''$.
The seeing (FWHM) was $\sim 0.9''$.

The modulus of the Fourier transform of the object
(visibility) was obtained with the speckle interferometry method
(Labeyrie 1970).
An image with a resolution of 320\,mas (Fig.~1)
was reconstructed from the
data set using the bispectrum speckle interferometry
method (Weigelt 1977; Weigelt \& Wirnitzer 1983; Lohmann et al. 1983;
Hofmann \& Weigelt 1986).

\section{Results}

\begin{figure*}
\includegraphics[width=10.6cm]{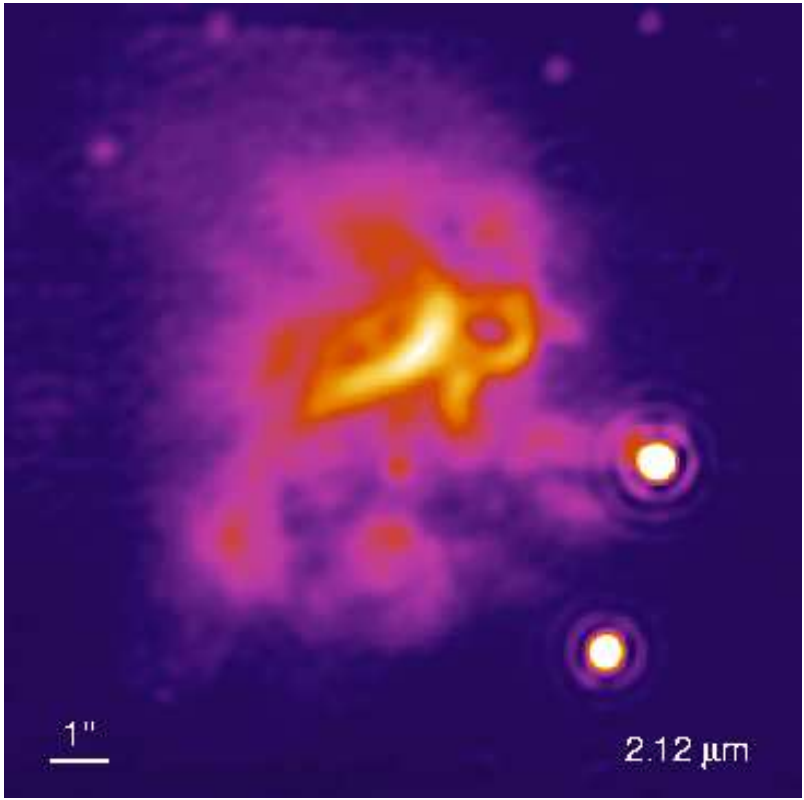}
\caption{Pseudocolor representation of our $K'$-band image of \I,
reconstructed using the bispectrum speckle interferometry method. 
The field of view is $\sim 11'' \times 11''$.
North is up, and east is to the left. We denote the upper of the two
bright point-like sources in the lower right image quadrant as IRS~1,
the lower one as IRS~2.}
\label{speckle_image}
\end{figure*}
\begin{figure*}
\includegraphics[width=10.6cm]{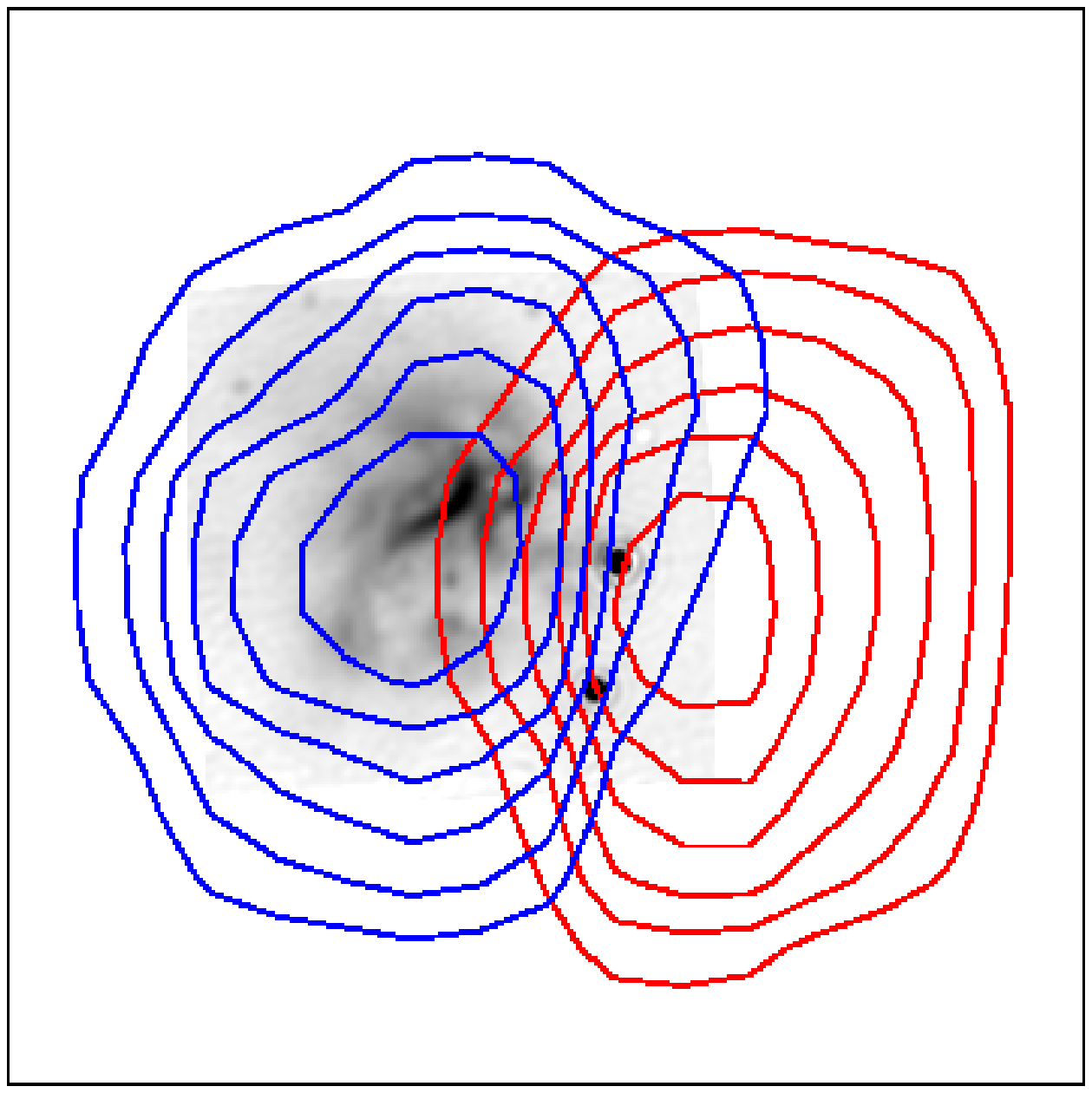}
 \caption{Comparison of our $K'$-band  speckle image of \I\ (greyscale image)
to the CO line maps from Beuther et al.~(2002b) shown as red and
blue contours for the red- and blue-shifted CO emission.
The field-of-view is $30''\times 30''$. }
   \label{composit}
    \end{figure*}

Our 320\,mas resolution speckle reconstruction of \I\ is presented in Fig.~\ref{speckle_image}.
Our image shows two point-like sources and extended diffuse nebulosity.
We denote the point-like sources as IRS~1 (the brighter, northern object)
 and IRS~2. The distance between IRS~1 and IRS~2 is $3.6''$.
Both objects appear unresolved at the resolution of our image.
The ratio of the fluxes of IRS~1 and IRS~2 is 1.53:1.

The total flux in our image is dominated by the diffuse nebulosity.
By measuring the fluxes of IRS\,1 and IRS\,2 in 
$0.8''$ diameter apertures,  we find that IRS\,1 contributes just 5.6\%
to the total flux in the image, and IRS\,2 only 3.7\%. This means that
90.7\% of the total flux in our $11'' \times 11''$ image
comes from the extended nebula.

Comparison of our speckle image with the images from
Hodapp (1994), the 2MASS images, and the 2MASS All-Sky Catalog of Point Sources
 suggests that IRS~1 corresponds to
2MASS~J23172102+5928480,
while the peak of the diffuse nebulosity corresponds to 
2MASS~J231721.57+5928496.
IRS~2 was not resolved as a separate point source in the 2MASS images.
A comparison to the $\sim 1.8''$ resolution seeing-limited images from 
Chen \& Yao (2004)
suggests that IRS~1 corresponds to their source NIRS~34, while IRS~2 
corresponds to their NIRS~35. Their source NIRS~19,
which they regarded as the brightest ``point source'' and thus most
massive ``star'' in \I, corresponds to the nebulous emission,
which is completely resolved in our image. The fact that
NIRS~19 is no stellar point source but apparently pure
nebulosity also explains the colors of $H-K' = 1.3$ and $J-H = 3.3$
derived by Chen \& Yao (2004),
which place the source in the ``forbidden'' region of the 
$H-K'$ versus $J-H$ color-color diagram (i.e.~to the left of the
reddening band for photospheric colors; see their Fig.~3).
These colors are consistent with the idea that 
a considerable fraction of the nebulous emission is scattered light.

\subsection{Structure of the diffuse nebulosity}

The nebulosity in the eastern part of our image shows a
remarkable wealth of detail.
Its structure is dominated by several curved features
which look like bow shocks. Such bow shocks are seen in many 
young jet and outflow sources and can usually be interpreted
as either the leading edge of jet-like streams of outflowing material 
or internal working surfaces within the jet flows.  
In remarkable contrast to ``usual'' bow-shocks, which are well aligned
and point away from the jet source, the
curved features in \I\ appear to be oriented in random directions. 

The nebulosity seems to be connected to IRS\,1, but not to 
IRS\,2. 
Considering the fainter parts of the nebula, as seen in the deep
seeing-limited $K$-band image in Hodapp (1994), the shape of the nebula
is clearly cone-like, with IRS\,1 at its tip. The axis of the cone
is along position angle PA $\sim 70\degr - 80\degr$ (similar to the PA of
$\sim 80\degr$ of the CO outflow), and the
opening angle is about $90\degr$. 
We also note that this
deeper image shows several extended ``finger''-like features
that also clearly point to IRS\,1; these features extend $\sim 20''$ outward
at position angles of $\sim 120\degr - 140\degr$.
IRS\,2 appears to be outside the cone-like
nebular structure and separated from the nebulosity.

Another very interesting feature is a narrow, elongated structure
that seems to  extend from IRS\,1 to the nebulosity.
The PA of this feature is $\sim 88\degr$. 
It displays a clumpy structure, with the brightest knot being located
$2.0''$ from IRS\,1. 
The width at this knot is $\sim 1.0''$ FWHM.

\subsection{Comparison of the near-infrared emission to the mm maps}

The detailed 
mm maps of \I\ presented by Beuther et al.~(2002b)
(beam size $11''$)
provide an opportunity to investigate the spatial
relation between the observed near-infrared emission,
the 1.2~mm emission, 
and the CO outflow structure.
For the alignment of the various images, we used the nominal telescope 
pointing position 
for the mm maps to establish the coordinate system.
Since the near-infrared source IRS\,1 can be identified with 
2MASS~J23172102+5928480,
we used its 2MASS position to align our speckle image with the
mm maps.
This alignment places IRS\,1 into the central peak of the 1.2~mm continuum
map
and just between the red- and blue-shifted CO map peaks.
In Fig.~\ref{composit} we have plotted the contours of the red- and blueshifted
mm emission over our $K'$-band speckle image.
The comparison suggests that the 
diffuse nebulosity coincides very well with the peak of the
blue-shifted CO emission.
We note that the astrometric uncertainties of this comparison are
significant. The positional uncertainty of the mm data (i.e.~the
possible pointing error) is $\leq 5''$, and due to the diffuse appearance
of the near-infrared sources in the 2MASS images, we estimate the
uncertainty of the 2MASS position of IRS\,1 to be $\sim 1''$.
Therefore, the formal uncertainty of our astrometric comparison does not
allow us to fully prove that IRS\,1 is actually in the center of the
mm peak.
However, for the reasons discussed below, we believe that IRS\,1 is 
actually well centered in the cloud core.

\section{Interpretation}

\subsection{Is IRS\,1 the massive protostar in \I?}

Many jet and outflow sources
are so deeply embedded in their surrounding molecular cloud material
that they are invisible in the near-infrared and can only be seen at longer
wavelengths. It is therefore not immediately clear that
 IRS\,1 is the massive protostar responsible for most of the infrared
emission in \I. It may
be an unrelated object, seen near the nebulosity by chance.
However, there are a number of arguments suggesting that IRS\,1 
is, in fact, the near-infrared counterpart of the massive protostar
driving the outflow:

First, we note that the 1.2~mm
emission is very compact with a size of $16.5'' \times 14.4''$
(FWHM, measured in the mm-map with $11''$ beam size). The
mm map, as well as all available mid- and far-infrared images, is consistent
with the presence of a single, unresolved point source.

Second, the recent high-resolution mid-infrared study of \I\
by Campbell et al.~(2003) suggested
the presence of a single unresolved point source. In their 
 $10\,\mu$m and $20\,\mu$m images with a spatial resolution of $\sim 1.5''$,
they found only a single point-like
source in the center of the mm peak and derived a FWHM of
$1.3''$ at $10\,\mu$m and $1.7''$ at $20\,\mu$m.
They also found that a major part of the mid-infrared emission comes from 
this point source.
These results clearly suggest a single dominant source in the molecular
cloud core, and it is very likely that the mid-infrared point source 
is identical to the massive protostar IRS\,1. IRS\,2 is not detected in these
mid-infrared images.

Third, the very close proximity (angular separation $< 2''$) 
of IRS~1 to the H$_2$O masers found
by Tofani et al.~(1995) also suggests that IRS~1 is the protostar driving
the outflow.

Finally, since
our comparison of the speckle image with the radio maps also
places IRS\,1 at the center of the mm peak, and since deeper near-infrared
images show no additional sources within $10''$ of IRS\,1 and IRS\,2,
we believe that IRS\,1
is identical to the mid-infrared source and, therefore, also 
to the massive protostar.

Further support for our assumption comes from the fact that IRS\,1 lies
at the tip of the cone-like nebulosity, and that 
the direction from IRS\,1 to the nebulosity (PA $\sim 73\degr$)
agrees  well with the CO outflow direction (PA $= 79\degr$).
The presence of the jet-like structure connecting IRS\,1 to the
diffuse nebulosity also suggest a physical relation.

\subsection{Relation of the near-infrared nebula to the outflow structure}

The positional coincidence of the point-like source IRS~1 with the center
of the mm emission and the diffuse near-infrared nebulosity with
the blue-shifted CO emission peak
suggests the following interpretation of the observed structures:
the bright, point-like $K$-band source is the protostar 
in the center of the compact molecular cloud core.
The protostar is embedded
in a dense circumstellar envelope or perhaps a thick circumstellar disk.
The outflow has cleared a cavity in the circumstellar
material, and what we see as the diffuse nebulosity
in our $K'$-band image is
a combination of light from the central protostar that is scattered
at the inner wall of this low-density outflow cavity or by material
within the cavity in our direction, and of $2.12\,\mu$m emission from
shock-excited molecular hydrogen in the outflow.
This interpretation is supported by the
$K$-band spectrum of the peak of the nebulosity  presented by
Chen \& Yao (2004), which shows strong emission in the $2.12\,\mu$m 
H$_2$ line.

The asymmetric general shape of our image, i.e.~the fact that we do not
see a counter lobe to the west of IRS~1,
 is easily explained as a geometrical effect.
The red-shifted outflow component has probably
cleared a similar cavity, which is, however, much fainter and essentially
invisible 
in our near-infrared image, because it is pointing away from us.
In fact, the deep $K$-band image from Hodapp (1994)
reveals a faint indication of the counter lobe to the west of IRS~1.
The continuum subtracted $2.12\,\mu$m H$_2$ line image
from Chen \& Yao (2004) also shows faint H$_2$ emission to the west of
IRS~1.


\subsection{Numerical simulations of a jet-driven outflow}

We attempt here to model the observed nebula features in \I\ more closely
and interpret them in the context of an outflow driven by a jet 
precessing through a wide angle. 
In particular, we try to explain the
prominent backward facing bow-shock like feature in the center of the 
nebulosity.
For this purpose we performed 
three-dimensional hydrodynamic simulations that take a heavy molecular jet
which cuts through a uniform external cloud, destroying some of the
molecules in its path.
In the fast precessing case, a reverse bow forms around the precession axis, 
from which the outflow is deflected (Rosen \& Smith 2004b).  This causes the outflow to be indented
along the flow axis. In contrast, we note that jets with
slow precession are shown to produce long curved streamers or multiple bow
shocks rather than reverse bows (Smith \& Rosen 2005).

Therefore, to determine if the observed strongly curved structures can be
better simulated, we revised the input parameters as follows.  We
suspected that the features observed in \I\ might correspond to an
outflow driven by a jet with an {\em intermediate precession rate}.
Clearly, we need a more specific definition of fast and slow
precession rates by comparing the rotation speed with the
outflow expansion speed. That is, the precession is fast if
the precession period is smaller than the outflow expansion time.
The latter is proportional to both the jet dynamical time scale $t_j = r_j/v_j$
and the ratio of jet to ambient density, $\eta = \rho_j/\rho_a$. It is also
inversely proportional to $\sin \theta$ where
$\theta$ is the half-angle of the precession cone, and $r_j$ and $v_j$ are
the initial jet radius and speed (Smith \& Rosen 2005).

Following published simulations, we took $v_j$ = 100~km~s$^{-1}$,
$r_j$ = 1.7\,$\times$\,10$^{15}$\,cm, $\theta = 20^\circ$ and $\eta = 10$.
This yields an outflow expansion timescale of $\eta t_j/ \sin \theta$ =
158\,yr. For comparison, published fast-precession simulations 
assumed a 50 year period while slow-precession runs took a 400 year period. 

\begin{figure}
\includegraphics[width=9.0cm]{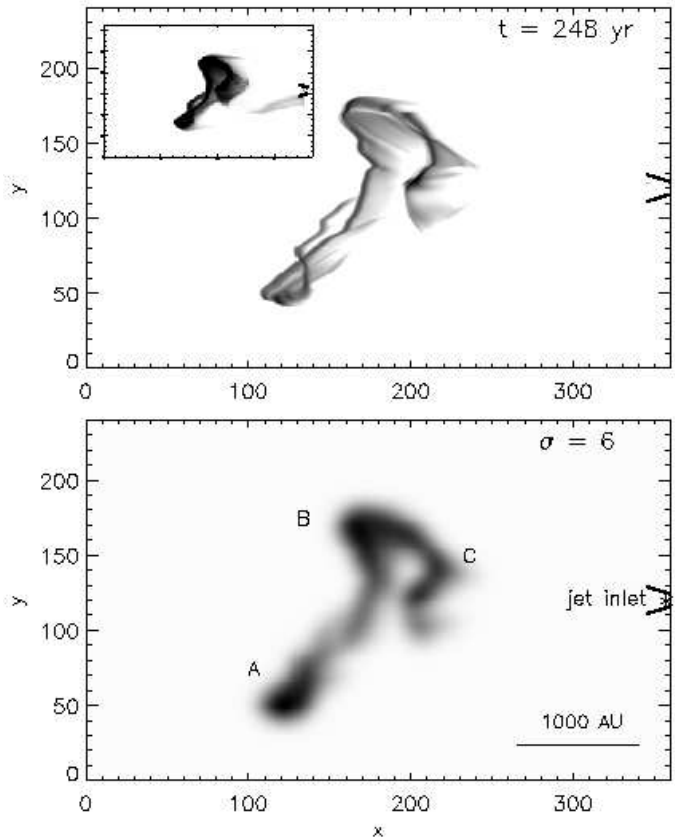}
 \caption{
Simulated images of the outflow resulting from a precessing
jet of intermediate precession rate. The  calculation was performed with a 
modified ZEUS-3D code on a grid of size $360\,\times\,240\,\times\,240$ 
zones. Each zone has a size of $2\,\times\,10^{14}$\,cm. Shown here are 
the 1-0 S(1) emission from H$_2$ at 2.12$\mu$m at the full numerical resolution
(upper panel) and  
convolved with a Gaussian with $\sigma$ = 6 zones (lower panel). The insert
shows the appearance of a wispy jet feature on an `over exposed' simulated
image. The ambient medium begins with 
a uniform density of $10^4$~cm$^{-3}$ and the jet possesses a density of  $10^5$~cm$^{-3}$
and a speed of 100~km~s$^{-1}$. The jet speed
was varied by superimposing ten per cent sinusoidal pulsations with a period of 60\,yr. 
No shear was applied. Full details of the
simulations can be found in Rosen \& Smith (2004ab). Note that the
view presented takes the axis about which the jet is precessing to be in the 
plane of the sky and the H$_2$ emission to be optically thin.}
\label{models_h2}
\end{figure}

Hence, we here took a 150 year period. The resulting H$_2$ simulation reproduced
the majority of the observed near-infrared features at the time of 248~yr after the
launch of the jet (Fig.~\ref{models_h2}). The brightness distribution, however, does not
entirely correspond, as discussed below.

\begin{figure}
\includegraphics[width=9.0cm]{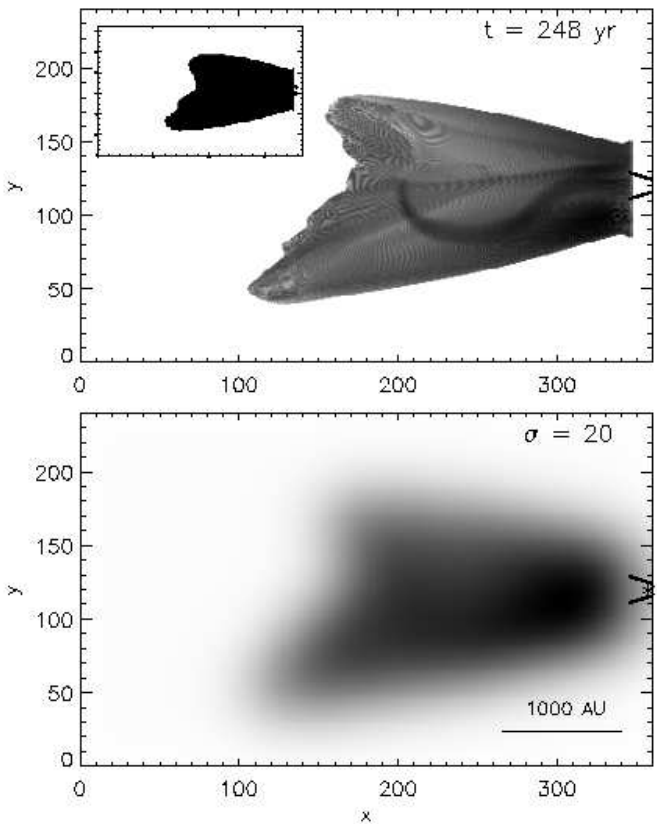}
 \caption{CO J=2-1 emission at 230\,GHz predicted from the simulation
presented in Fig.~\ref{models_h2}. Only the CO which has been entrained
(i.e. set in motion) is included in the image (i.e. the background cloud
has been subtracted). The lower panel demonstrates the
structure which would be detected when observed at low spatial resolution.
The inset panel is an over-exposed simulation.
\label{models_CO}}
\end{figure}

Note that in the figure, the jet enters from a precessing nozzle
on the right of the panel. The leading edge of the reverse bow (marked as A) has reached
a distance of  5\,$\times$\,10$^{16}$\,cm from the nozzle. In the
observation, the edge of the bow is $\sim 6\arcsec$ from the driving star. At
a distance of 5.7\,kpc, this is an extent of   5\,$\times$\,10$^{17}$\,cm,
considerably larger than in the simulations. This implies that the simulation time scales should be
scaled by a factor of 10 upwards, assuming a similar jet speed, to give a precession
period of 1,500 yr and a source age of 2,500 yr. 
The lower panel of Fig.~\ref{models_h2} displays the molecular hydrogen image convolved 
with a Gaussian beam with a standard deviation of 6. When scaled up to the observations of the source, 
this corresponds to about $0.15''$ and produces structure close to that of the speckle
image.

How can  this specific feature, so  similar to the one observed, be found associated with a jet flow into a 
uniform density medium? Part of the answer lies in the intermediate precession period which allows the jet material 
to sweep out an arc-shaped shock into the ambient medium rather than a bow shock. However, although the jet is dense,
the change in direction implies that the jet momentum is spread out over the same arc. Consequently, the ambient medium
is able to provide resistance to the advancing jet and distort the shape of the arc. In addition, ten per cent perturbations
to the magnitude of the jet speed have been superimposed which accelerate the distortion of the filamentary shock.

The simulated CO structure for the same execution is displayed in Fig.~\ref{models_CO}.
Note that the leading edge is indented even when convolved,
as seen in the observations,  although only just perceptible. However, all other structure is
smeared out. The dynamical time derived from the CO
data is a few times that derived for the precessing H$_2$ feature
currently prominent. The true age of the bipolar outflow now being observed may thus be
much shorter, being driven by fast jets. However, it also seems
plausible that a massive outflow has been in existence for a time
exceeding 10,000~yr to be consistent with their apparent ubiquity.
This suggests that massive stars may also undergo outburst episodes,
possibly related to FU Ori type accretion events or merging events.

The observed $K$-band arcs of emission could arise from 
vibrationally-excited H$_2$, scattered continuum emission from stellar object,
or emission from warm dust grains.
Most likely, the observed emission is a combination of all three processes.
In the numerical study, the molecular hydrogen is vibrationally excited only in shocks, being calculated
from formula which take into account collisional excitation and collisional and radiative de-excitation.
The shocks from which the near-infrared emission originates are J-type of speed exceeding $\sim$ 8~km~s$^{-1}$,
capable of heating the gas above $\sim$ 1000\,K in gas of density exceeding $10^4$~cm$^{-3}$.
As a consequence, along with distortion and projection effects of the shock front,
the strongest emission is predicted to occur from the
edges of the rim structures (locations A and B in Fig.~\ref{models_h2}) whereas the observed emission appears to
peak towards the center (the flow axis). The emission from nearer the source, within region C in the simulated image,
may also be influenced by stronger extinction.
Fluorescent excitation of H$_2$
and scattering of radiation from the central source thus seems to dominate the 
observed emission around the flow axis. 
Narrow-band imaging and polarimetry
would be helpful to distinguish between these possibilities.

The observed backward facing curved shock can be understood as a 
remnant of the
initial ring of shocked gas around an inward pointed cone of ambient
material trapped within the volume of material swept up by the precessing jet.
The ring of shocked gas fragments into numerous bow-shocks and curved
filaments, but the  head of the inward pointed cone remains visible
for some time and points towards the jet source. 

Further support for our interpretation of a jet-driven flow 
comes from the observed elongated  feature connecting IRS\,1  to the nebulosity.
This feature may trace of the most recently ejected material
in the precessing jet. It
points just towards the southern edge of the most
prominent bow-shock like structure, consistent with the idea
that the bow represents an inward-pointed cone that has been created
by a jet precessing around the cone's surface.
The jet of emission from the driving source is not visible in our simulated
image, what
may well be due to the absence of short period oscillations in the jet.
In any case, pulsations tend to decay very rapidly within jets and do not
readily produce elongated jet structures.
Nevertheless, by changing the cuts in our data to enhance weaker features
(insert of Fig.~\ref{models_h2}),
we do indeed find a jet-like structure in the observed direction. It appears to be related to the
transverse motion of the precessing jet, which pushes into relatively
undisturbed cavity material. In other words, a continuous jet-like
structure can be produced due to the interaction between a wandering jet
and the cavity walls.

In summary, modeling suggests that the outflow is jet-driven and the jet 
precesses with a period of $\sim$ 1500~yr. Such a precession period may be
triggered by a companion protostar either orbiting within 30--100~AU (i.e. 
within a 10~M$_\odot$ system with a period of 50--300\,yr)
and/or in the process of merging.

\section{Comparison of \I\, with other intermediate- and 
high-mass YSOs and radiative transfer model images}
\label{comparison}

\begin{figure*}
\includegraphics[width=18cm]{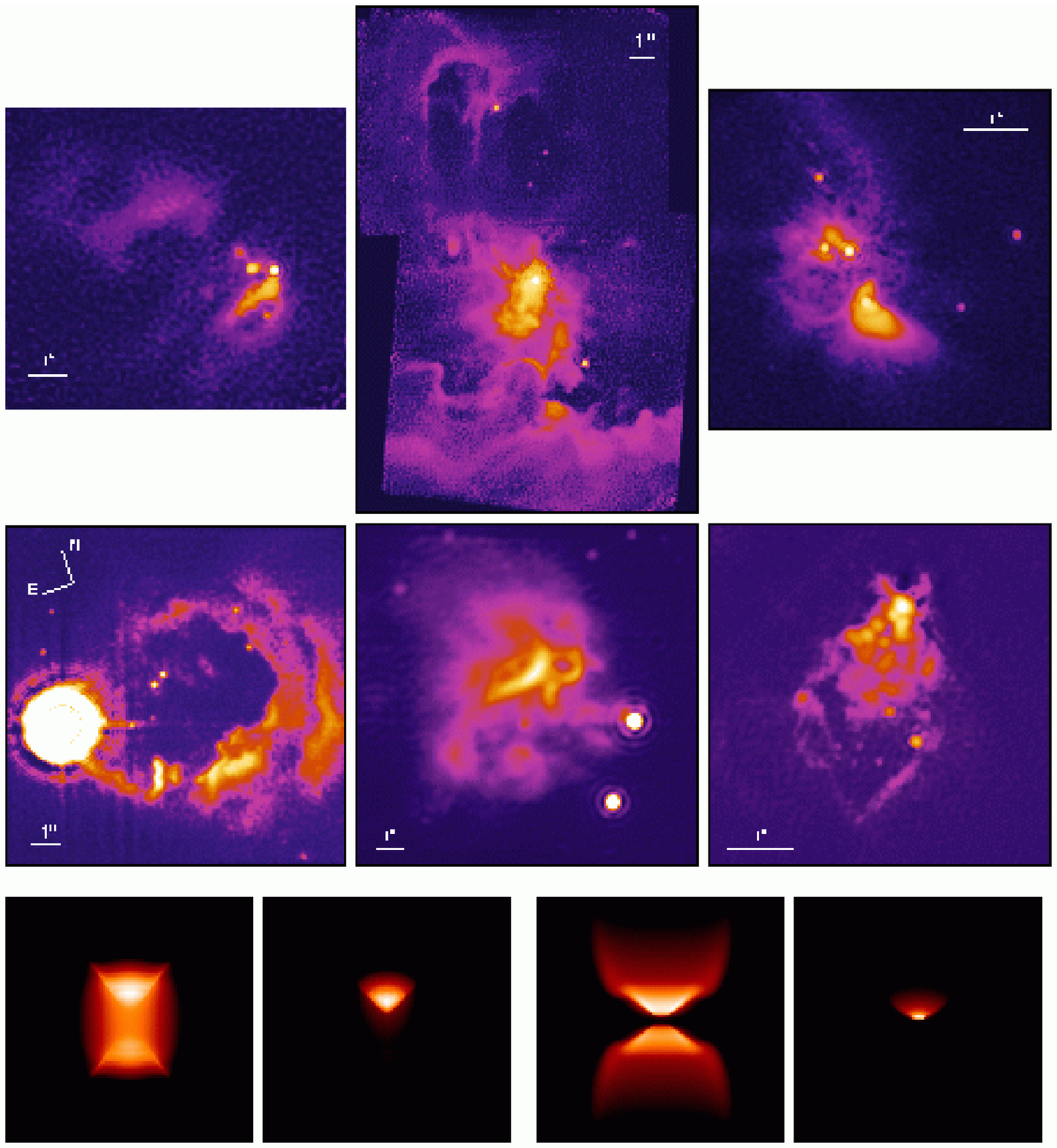}
 \caption{Comparison of $K'$-band images from bispectrum speckle interferometry 
of several intermediate- and high-mass YSOs to radiation transfer simulations.
\newline
Upper two rows: Pseudocolor representations of several images. 
The objects are
arranged by increasing estimated luminosity (or mass) of the YSO.
The first row displays S140~IRS3, S140~IRS1, and Mon~R2~IRS~3 (from left to
right).
The second row shows AFGL~2591, \I, and K3-50~A.
For details and references see text.
In all images north is up, and east is to the left.
The length scales in AU corresponding to the one arcsecond bars shown in the
images are given by the numbers in Table 1 col.~4.
Lower row: simulated $K$-band images derived from 2D radiative transfer 
calculations. 
The two left images correspond to a spherical envelope model with cone-like
cavities (Model A), seen under inclination angles of $i=85\degr$ and $i=60\degr$. The
two right images are from a model of a Keplerian disk embedded in
a spherical halo with polar outflow cavities (Model B), again for 
inclination angles of $i=85\degr$ and $i=60\degr$.
Each image shows a $1250~{\rm AU} \times 1250~{\rm AU}$ region.
 For details of the models
see the text.
}
\label{mosaic}
\end{figure*}

Complex structures are generally detected in the immediate 
circumstellar environments of intermediate- and high-mass protostars,
and \I\, is no exception. 
In Fig.~\ref{mosaic} we show a mosaic of $K$-band bispectrum-speckle
images from several intermediate- and high-mass protostars
that have been observed by our group during the last few years.
As illustrated in Fig.~\ref{mosaic},
different objects display intriguingly different near-infrared morphologies.
This is, perhaps, not surprising given the following vital intrinsic sources of 
influence:
\begin{itemize}
\item the evolving properties of the central radiating source,
\item the varying impact and disruption by the jet and/or wind,
\item the clumpy small-scale distribution of surrounding material, and
\item the configuration of large-scale geometric components.
\end{itemize}
In addition, the viewing angle is an extrinsic factor which needs to be 
taken into account
before we can relate the structure to the physical cause.  With this purpose,
we here compare bispectrum speckle images to each other and to radiative transfer
simulations.

\subsection{Observed infrared morphologies}

The relevant properties for the six objects
shown in Fig.~\ref{mosaic} are listed in Table~\ref{thesources}.
The major qualitative features are summarised below.

\begin{table}
 \caption{Basic data for the six sources discussed in \S~\ref{comparison}. Note that 
there is a strong dependence on the poorly known distances. The length scales in AU
corresponding to the one arcsecond scale, as indicated by the bars shown in  Fig.~\ref{mosaic}, 
are equivalent to the distance in parsecs (column 4).}
 \label{thesources}
 \begin{tabular}{lrrrr}
 \hline
 \noalign{\smallskip}
  Object     &Luminosity   & ~~~Mass      &~~~~Distance     &  \\
              &$[L_\odot]$   &$[M_\odot]$ &[pc]   \\
 \noalign{\smallskip}
 \hline
 \noalign{\smallskip}
 {S140~IRS3}         & $3 \times 10^3$ &  ~~8    &   1000   &       \\
 {S140~IRS1}         & $5 \times 10^3$ &  ~~10  &   1000   &       \\
 {Mon~R2~IRS\,3}     & $1 \times 10^4$ &  ~~10   &   800 &      \\
 {AFGL~2591}         & $2 \times 10^4$ &  ~~12   &   1000   &  \\ 
 {IRAS~23151}        & $1 \times 10^5$ &  ~~25   &  5700  &  \\
 {K3-50A}            & $2 \times 10^6$ &  ~~30   &  8700  &   \\
\noalign{\smallskip}
 \hline
 \end{tabular}
 \end{table}

{\bf S140 IRS3} is resolved into three point sources and
an extended diffuse feature north-east of IRS\,3 that
displays a remarkable $\sf S$-shaped structure (Preibisch et al.~2001). 
This feature is the innermost part of an at least
$15''$ long structure that extends towards a bow-shock like
patch located $90''$ away from IRS\,3. 
It is also associated with strong, shock-induced H$_2$ line emission,
indicating a collisional interaction between outflowing material and the ambient medium.
The $\sf S$-shaped structure was successfully reproduced by a model involving a
precessing outflow stemming from IRS\,3A (the brightest point source in the 
image).

The brightest feature in the image of {\bf S140 IRS1}
is the bright, extended, and very clumpy structure
pointing away from the central source, IRS1, towards the south-east,
in the same direction as the
blue-shifted CO outflow lobe (Schertl et al.~2000, Weigelt et al.~2002a).
We interpret this feature as the clumpy inner surface of a partially
evacuated cavity  which has been excavated by the strong outflow from IRS1.
A system of three arc-like structures to the north-east
could be produced by a precessing jet interacting with cavity walls.
The systematic positional offsets of their bow-shaped tips
then correspond to the successive sites of jet impact. These results
provide direct evidence for the existence of two distinct bipolar outflow
systems driven simultaneously from IRS\,1.

The speckle images of {\bf Mon~R2~IRS\,3}
reveal a close triple system surrounded by 
diffuse nebulosity (Preibisch et al.~2002).
A prominent bipolar nebula surrounds IRS\,3~A (the brightest point source
just beneath the center of the image). 
IRS\,3~B (the second brightest point source, just above the center of the 
image) shows a remarkable jet-like emission feature pointing
towards the north-east.

The {\bf AFGL~2591} image displays a well-defined conical
structure with a bright infrared source
at the apex (Preibisch et al.~2003). 
Several loops of nebulosity are clearly seen, which
suggest that shock waves driven by discrete outbursts 
have swept through the cavity.

The image of the ultracompact H~II region {\bf K3-50A} resolves the central 
$1''\times 1''$ region into at least 7 point-like objects (Hofmann et al.~2004). 
The cone-shaped nebulosity extending to the south contains
considerable fine-structure.
The brightest $K'$-band source is located exactly at the tip of the
cone. The nebula also shows several arcs.
The orientation of its main axis corresponds to the direction
of the CO outflow from  K3-50A.

\subsection{Radiative transfer simulations}

We have performed a number of radiation transfer simulations of
YSOs surrounded by envelopes and/or disks in order to better
recognize the physical structure behind the features in the speckle images.
We applied the 2D radiation transfer code described 
in Sonnhalter et al.~(1995).  In this code, 
the distribution of dust temperatures and radiation intensities in
an axially symmetric, dusty circumstellar environment around a central
radiation source is calculated within 
the framework of the flux-limited diffusion approximation (Levermore \&
Pomraning 1981).

The equations are discretized on a quadratic, equally spaced grid.
To improve resolution and convergence, a system of 5 nested grids with
decreasing grid spacing is used. The differential
equations for the radiation field are iterated together with the temperature
equations until a self-consistent equilibrium
configuration is reached.
After the determination of the dust temperatures for a 
multi-component dust model,
a ray-tracing procedure is used to calculate intensity maps
for the appearance of the
central object and its circumstellar environment at varying inclinations
for selected frequencies.
A detailed description of the code is given in Sonnhalter et al.~(1995). 

In our simulations we assume a luminosity of $20\,000\,L_\odot$ and an
effective temperature of 30\,000~K for the central source.
The dust model is described in the Appendix.
It was derived  from Preibisch et al.~(1993), and was constructed
to reproduce the dust properties in dense molecular clouds. 

We consider two different models for the density distribution of the 
circumstellar material. Model~A assumes that the central source 
is embedded in a dusty {\it envelope} with a radial power-law density distribution
that is bounded by a Fermi-type function at an outer radius of
$3.5 \times 10^{15}$~cm (233~AU).
The cavities are simulated as a pair of cones
with their tips at the position of the central star, in which the
density is 1000 times smaller than in the other parts of the envelope at the
same radial distance from the center.
The model details are provided in the Appendix.

Model B assumes a geometrically thick Keplerian {\it disk},
which is embedded in a spherical halo that has wind-blown cavities along the
rotational axis. We employ an analytical
model for a Keplerian disk, which is expected if
the central potential of the star is the dominating gravitational
potential in the disk. Again, the details are provided in the Appendix.

Simulated $K$-band images for the two models with different inclination angles
are shown in the lower row of Fig.~\ref{mosaic}. The contrast range 
displayed in the images is 4 orders of magnitude. 
For high inclination angles ($i=85\degr$, i.e.~nearly edge on) 
the direct view towards
the central source is blocked, and the dominant features in the
simulated images are scattering lobes above and below the disk or torus
plane. 
For low inclination angles ($i=60\degr$), the observer is able to ``look through''
the less dense regions of the cavities towards the hottest and, therefore,
bright central regions. Hence, the extension of the scattered light
is relatively small for a given contrast ratio.
Note that these model images should only be compared to the 
large scale morphology in the observed images. The small-scale
clumpiness and bow-shocks captured in
the speckle images cannot be simulated
in our 2D radiative transfer model.

\subsection{Comparison of observed morphologies and simulations}

A common feature in the menagerie of images is
diffuse emission extending from the central point-like source with a
fan-shaped morphology.
The detailed shape and structure of the diffuse emission
varies strongly from object to object.
A critical factor in the appearance of the structures
is the inclination between the outflow/cavity axis
and the line-of-sight.
In fact, the radiation transfer simulations described below
demonstrate that the cone of the bipolar cavity oriented towards us appears very
 bright
for moderate or low inclination angles ($i \la 75\degr$).
The opposite cone is considerably fainter or completely 
invisible because it is oriented away from us and
hidden by circumstellar extinction. 
Objects with low inclination angles are thus suggested to be
S140~IRS~1, K3-50A, and \I.
In systems with high inclination angles ($i \ga 75\degr$),
a more symmetric morphology is expected.
A good example in our sample is Mon\,R2~IRS\,3A where, in addition to
the bright nebulosity toward the south, fainter nebulosity
can be seen north of the point source, resembling a bipolar reflection nebulosity.

The observations indicate that a second difference is related to
the intrinsic asymmetry of the cavity
structure itself. In some objects the shape of the cavity appears
highly symmetric (e.g.,~S140~IRS1, AFGL~2591,
K3-50A), while other objects show remarkable asymmetry (e.g.,~S140~IRS3).

A third major difference is the relative amount of reflected light and 
$H_2$ shock
emission. In S140~IRS1 and Mon\,R2~IRS\,3A, the infrared 
morphology is strongly dominated by 
light reflected from the inner cavity walls.
In other objects we observe emission from shocked material in the vicinity of
the jets (e.g., the north-eastern jet from
S140~IRS3;  see Preibisch et al.~2001) or outflowing shock waves seen in 
reflected light, such as
the loops in AFGL~2591.

Can we distinguish a jet-driven outflow model from a wind-driven model?
The jet-driven outflow model predicts highly collimated axial flows,
while the wind-driven model predicts wide-angle
shells and loops.
Many of the YSOs in our sample suggest
both kinds of outflow
activities. 
The stellar wind is expected to play an
increasingly important role in driving the outflows
with increasing mass of the YSO.
Furthermore, a wide-angle wind is expected to clear a broader path in
the surrounding circumstellar matter than a highly collimated jet.
In our sample of intermediate-to-high mass YSOs we see some indication
that the cavity opening angle increases with increasing luminosity
of the central YSO, tentatively supporting this suggestion.
However, jet precession and multiple outflows also clearly
act to broaden cavities in the two lower mass examples, making any
relationship less obvious.

Comparing the observed images to the radiative transfer 
simulations, we find that
the objects showing triangular-shaped morphologies with rather straight walls
(AFGL~2591, \I, and K~3-50\,A; second row in Fig.~5) 
are quite well reproduced with
Model A, the inclined
``envelope with cavity'' model.
For S140\,IRS3 and S140\,IRS1 (upper row in Fig.~5),
the very elongated nebulosities 
may indicate a very small cavity opening angle.
The observed bipolar structure around Mon\,R2\,IRS\,3\,A (Fig.~5 c), 
on the other hand, appears reminiscent of the disk model B, seen under a
rather high inclination angle. 

To summarize,  the large-scale features observed in the 
speckle images can be understood in the frame of simple disk or
envelope plus cavity models, whereas the small scale structure depends
on additional factors such as clumpiness of the circumstellar material
and the presence of shock waves.

\section{Summary and Conclusions}

A bispectrum speckle image of the massive protostellar object 
\I\ reveals complex structures with several arcs in 
the infrared nebulosity to the north-east of the driving source.
The nebulosity is coincident with
the blue-shifted CO outflow, which is consistent with a model 
in which the red-shifted
lobe of the outflow is hidden from view in the near-IR through intervening
extinction.
The general morphology of the arcs in the nebulosity can be reproduced by
a model assuming that the outflow is driven by a precessing jet.
Our findings constitute another example for an apparently jet-driven 
outflow from a massive protostar and support the assumption 
of a common mechanism for the formation of outflows from protostars 
of all masses.
We have performed new simulations of 3D hydrodynamic precessing jets
and 2D radiative transfer in cavities. On comparison to a set of speckle images,
we conclude that the inner infrared
structures of high-mass stars are strongly influenced by the
orientation and symmetry of the bipolar cavity, as well as the relative
strength of shocked and reflected emission components.

Finally, we can speculate about the reason for the
precession of the outflow from \I.
One possible explanation might be that the outflow source is a member 
of a close (unresolved in our images) binary system
 where the rotational axis of the star driving
the outflow is misaligned with the orbital plane of the binary
(see, e.g., Terquem et al.~1999; Bate et al.~2000).
Other possibilities include anisotropic accretion events that altered
the angular momentum vector of the disk, or
radiation-induced warping of the disk, as discussed by Shepherd et al.~(2000).

\acknowledgements{We would like to thank
the MMT staff for their support of this
run, D.~Apai and I.~Pascucci for assistance with the observations, and
the referee for a very  careful report that helped to improve the
paper.
H.B.~acknowledges financial support by the
Emmy-Noether-Program of the Deutsche Forschungsgemeinschaft (DFG,
grant BE2578/1).
M.R.M.~acknowledges support from the Cottrell Scholars
Program of the Research Corporation.
M.D.S.~thanks Alex Rosen for code development,
and DCAL, PPARC and INTAS for funding.
This publication makes use of data products from the Two Micron All Sky Survey, which is a joint project of the University of Massachusetts and the Infrared Processing and Analysis Center/California Institute of Technology, funded by the National Aeronautics and Space Administration and the National Science Foundation.
This research has made use of the VizieR catalogue access tool, CDS, Strasbourg, France}

\appendix
\section{The radiative transfer model}\label{appx}

The dust model consists of small
amorphous carbon grains and large silicate grains and assumes that the
silicate grains are coated with a mantle of ``dirty ice" if their
temperature is below 125~K.
The radii $a$  of the grains obey a power law  distribution of the
form $  n(a) \propto  a^{-3.5}$,
where we assume that the carbon grain
radii range from 7 nm to 30 nm and the silicate radii from 40
nm to 1~$\mu$m. The ratio between the radii of ice-coated and uncoated
silicate grains is $1.145$. The ``dirty ice" consists of a 3:1
(volume) mixture of H$_2$O and NH$_3$, which is polluted with 10~\%
carbon particles. When the ice coatings sublimate, the carbon particles
are set free
and a ``naked" silicate grain remains. The sublimation temperatures of
the silicate and carbon particles are taken to be 1500~K.
The destruction of grains through
sublimation is simulated by setting the number density of the
grains to zero, if the temperature of a particular grain component
calculated simultaneously with the radiation field is above the
sublimation temperature.
The inner boundary of the dusty environment around the central source
is therefore given by the point of dust
sublimation.

The envelope density in Model~A is
\begin{equation}
\rho_{\rm envelope} =   \left\{ \begin{array}{lcrcr}
 \phantom{\frac{1}{1000}\times\,\,} \rho_{\rm halo}  & \mbox{for} &z/r&< & 1.5\\
   \frac{1}{1000}\times \rho_{\rm halo}  & \mbox{for} &z/r&\ge  & 1.5\end{array}\right.
 \label{den_torusII}
\end{equation}
with
\begin{equation} 
\rho_{\rm halo} = \rho_0 \times \left(\frac{\D r}{\D r_0}\right)^{-1.25}
\times \left (1+e^{\D \left (r-r_0 \right )/\alpha}\right )^{-1}
\end{equation}
where $\rho_0$ is the density of the envelope at a characteristic radius
$r_0 = 3.5 \times 10^{15}$~cm = 233~AU and $\alpha = 0.12 \times r_0$.
The full opening angle of the cavities is $67\degr$.

The total density in the disk Model B can be written as
\begin{eqnarray}
\rrho(z,r)&=&
\rrho_{\rm disk}\left ( 1+e^{\D\left (r-r_s\right )/\alpha}\right )^{-1}\\
&+&\rrho_{\rm halo}\left [\left (1+e^{\D-\left (r-r_s \right
  )/\alpha}\right ) \right. \nonumber\\
&&\times \left . 
\left (1+e^{\D\left (z-z_h \right )\cos\theta/\alpha}\right)\right]^{-1} \!\!\!\!\!\!. \nonumber
\end{eqnarray}
The parameters $r_s$ and $z_s=h\, r_s$ fix the radius and height of the
disk while  $\alpha$ determines the width of the transition zone of the
Fermi-type functions. The halo density $\rrho_{\rm halo}$ is given by
\beq
\rrho_{\rm halo}=\rrho_s\frac{r_s^2}{r^2+z^2},
\eeq
and $\rrho_s$ is the density of the disk at $r=r_s$ and $z=0$.  Points
with $z > z_h$ lie within the cavities. They are modeled by

\beq
z_h=z_s \left ( \frac{r}{r_s} \right )^4,
\eeq
The slope of $z_h$ fixes $\theta$ by
\beq
\cos\theta=\left ( 1+\left (4\,\frac{z_h}{r} \right )^2 \right )^{-1/2}.
\eeq
For the density of the disk $\rrho_{\rm disk}$  we take an analytical
model for a Keplerian disk, assuming that
the central potential of the star is the dominating gravitational
potential in the disk:
\beq
\rrho_{\rm disk}=\rrho_s \left(\frac{r_s}{r} \right )^{15/8}
e^{\D -\pi/4
\left (z/z_s\left(r_s/r \right )^{9/8}\right )^2}.
\eeq
We fixed the parameters to $r_s = 3.5 \times 10^{15}$~cm = 233~AU,
$h = 0.4$, and $\alpha = 0.1 \times r_0$.

\end{document}